\documentclass[sigconf, nonacm, screen]{acmart}

\usepackage{algorithmic}
\usepackage{array}
\usepackage {booktabs}
\usepackage{url}
\usepackage{multirow}

\newcommand{\fblas}{{\footnotesize F}BLAS}
\newcommand{\POSIT}{Posit(32,2)}
\newcommand{\POSITA}{Posit(32,2)\ arithmetic}
\newcommand{\Rgemm}{{\it Rgemm}}
\newcommand{\Rgetrf}{{\it Rgetrf}}
\newcommand{\Rgetrs}{{\it Rgetrs}}
\newcommand{\Rpotrf}{{\it Rpotrf}}
\newcommand{\Rpotrs}{{\it Rpotrs}}
\newcommand{\etal}{\textit{et al.}}

\newcommand\blfootnote[1]{
    \begingroup
    \renewcommand\thefootnote{}\footnote{#1}
    \addtocounter{footnote}{-1}
    \endgroup
}

\copyrightyear{2024}
\acmYear{2024}
\setcopyright{acmlicensed}\acmConference[HPCAsia 2024]{International Conference on High Performance Computing in Asia-Pacific Region}{January 25--27, 2024}{Nagoya, Japan}
\acmBooktitle{International Conference on High Performance Computing in Asia-Pacific Region (HPCAsia 2024), January 25--27, 2024, Nagoya, Japan}
\acmPrice{15.00}
\acmDOI{10.1145/3635035.3635046}
\acmISBN{979-8-4007-0889-3/24/01}

\begin{document}
\title{Evaluation of POSIT Arithmetic with Accelerators}


\acmConference[HPC Asia 2024]{The International Conference on High Performance Computing in Asia-Pacific Region}{January 25--27, 2024}{Nagoya, Japan}

\author{Naohito Nakasato}
\affiliation{
\institution{University of Aizu}
\city{Aizuwakamatsu}
\state{Fukushima}
\postcode{965-8580}
\country{Japan}
}
\email{nakasato@u-aizu.ac.jp}

\author{Yuki Murakami}
\affiliation{
\institution{National Institute of Information and Communications Technology}
\city{Koganei}
\state{Tokyo}
\postcode{184-8795}
\country{Japan}
}

\author{Fumiya Kono}
\affiliation{
\institution{Shizuoka Institute of Science and Technology}
\city{Fukuroi}
\state{Shizuoka}
\postcode{473-8555}
\country{Japan}
}

\author{Maho Nakata}
\affiliation{
\institution{RIKEN}
\city{Wako}
\state{Saitama}
\postcode{351-0198}
\country{Japan}
}

\begin{abstract}
We present an evaluation of 32-bit POSIT arithmetic through its implementation as accelerators on FPGAs and GPUs.
POSIT, a floating-point number format, adaptively changes the size of its fractional part.
We developed hardware designs for FPGAs and software for GPUs to accelerate linear algebra operations using \POSITA{}.
Our FPGA- and GPU-based accelerators in \POSITA{} significantly accelerated the Cholesky and LU decomposition algorithms for dense matrices.
In terms of numerical accuracy, \POSITA{} is approximately 0.5 - 1.0 digits more accurate than the standard 32-bit format, especially when the norm of the elements of the input matrix is close to 1.
Evaluating power consumption, we observed that the power efficiency of the accelerators ranged between 0.043 - 0.076 Gflops/watts for the LU decomposition in \POSITA{}.
The power efficiency of the latest GPUs as accelerators of \POSITA{} is better than that of the evaluated FPGA chip.
\end{abstract}

\maketitle

\blfootnote{\copyright Owner/Author 2024. This is the author's version of the work. It is posted here for your personal use. Not for redistribution. The definitive Version of Record was published in HPCAsia '24: Proceedings of the International Conference on High Performance Computing in Asia-Pacific Region, January 2024, Pages 62-72 \url{https://dx.doi.org/10.1145/3635035.3635046}.}

\section{Introduction}
Floating-point arithmetic is a fundamental tool for solving problems in computational science.
To this end, we use 64-bit double-precision (DP; IEEE binary64) floating-point (FP) arithmetic as the conventional way to implement numerical computations and simulations. 
For instance, in large-scale parallel computer systems or high-performance computing (HPC), one key performance metric of computer systems is how fast a system solves dense linear equations using DP arithmetic operations commonly referred to as LINPACK. 
The consensus among computational scientists is that the DP arithmetic has enough precision, e.g., $\sim$ 15 digits, for most applications. 

The past decade has seen significant advancements in machine learning (ML) algorithms, particularly in neural networks (NN), and their accelerated implementation using GPU-based parallel computing systems and domain-specific architectures (DSAs) for ML algorithms. 
These GPUs and DSAs are used to train large language models (LLM) with over 175 billion parameters \cite{2020arXiv200514165B}.
To enable such training of large NN models on a massive scale, standard 64-bit DP arithmetic is not always necessary.
Therefore, a new area of research is now devoted to optimizing the appropriate arithmetic format according to the required arithmetic precision for ML algorithms.
For this purpose, arithmetic formats with much shorter data lengths than the DP format were proposed, such as binary16 (IEEE 754R) and fp8/fp9 \cite{8344479}, as well as new FP arithmetic formats, such as BrainFloat-16 and TensorFloat-32.
These new formats have been implemented in existing GPUs (e.g., NVIDIA A100/H100 and AMD MI200) and DSAs (e.g., Google TPU, Graphcore M-2000, Cerebras CS-2, SambaNova Cardinal SN30). 


The POSIT arithmetic, a component of the newly proposed Universal Number (Unum) FP arithmetic by Gustafson \cite{gustafson2017beating}, addresses several issues present in standard FP arithmetic formats, such as IEEE binary32 and binary64.
Specifically, Unum type III, or POSIT, is a hardware-friendly FP format in which the bit length allocated to the fractional and exponent parts is variable and depends on the magnitude of an FP number.
This design allows for a fixed total bit length but dynamically allocates more bits to the fractional part when a number is close to 1.
This feature is particularly well-suited for applications in computational science, where partial differential equations are solved in non-dimensional units.
Additionally, the POSIT format holds the potential for effectively dealing with NN models because we need stabilization techniques for their training.
Specifically, we stabilize the training by enforcing the outputs from NN layers close to 1 through normalization techniques employed in the middle of the layers \cite{2015arXiv150203167I} or by structuring NN layers in the residual form \cite{7780459}.

In this study, we assess the performance and power efficiency of the POSIT arithmetic as implemented on GPUs and Field Programmable Gate Arrays.
To establish a baseline, we use SoftPosit \cite{softposit}, a C language software library, as our reference.
While many POSIT implementations are available as software libraries, they have yet to be evaluated on GPUs.
One reason for this is that modern datacenter GPUs, such as NVIDIA A100 or H100, have primarily focused on providing faster FP performance in binary64 for traditional HPC applications until recently.
However, modern GPU architectures are also highly effective for integer operations since their compute components, such as CUDA cores, can execute both FP and non-FP (integer) instructions.
In response to the growing demand for GPUs in ML and data science, recent trends have led to the inclusion of more non-FP instructions, such as Dynamic Programming X instructions in CUDA 12.
We found that porting a portion of SoftPosit as GPU kernels proved to be a highly effective means of evaluating the POSIT arithmetic, as its execution time on GPUs is significantly faster than on CPUs.

The application of reconfigurable computing in implementing and evaluating a novel FP arithmetic is typically achieved through FPGAs.
Shortly after the introduction of the POSIT format in 2017, POSIT implementations for FPGAs were presented in several works, including \cite{podobas2018posit} and \cite{jaiswal2019pacogen, hou2019enhancing, montero2019template, 9180771, 8615707, FloPosit, 8892116, forget:hal-03195756}.
In \cite{podobas2018posit}, the authors demonstrated vector-vector multiplication using 32-bit POSIT arithmetic.
Our work builds upon these works by implementing an accelerator for POSIT arithmetic that specifically targets linear algebra operations, such as matrix-matrix multiplication.
Additionally, our work is complementary to \cite{9139786}, in which POSIT arithmetic was applied to the Conjugate Gradient method and Cholesky decomposition for sparse matrices as iterative solvers.
In this study, we directly compare the performance of our FPGA designs with our GPU implementation by solving two decomposition algorithms for dense matrices: Cholesky and LU decompositions.

The main contributions of our work are as follows:

\begin{itemize}
\item We conducted a comprehensive evaluation and comparison of 32-bit POSIT arithmetic on FPGAs and GPUs.
\item We assessed the numerical error of the Cholesky and LU decomposition algorithms when utilizing POSIT arithmetic.
\item We examined the power efficiency of our POSIT arithmetic accelerators specifically for the LU decomposition.
\end{itemize}

This paper is structured as follows: Section \ref{sec_posit} provides a brief overview the POSIT format.
Section \ref{sec_acc} offers a detailed description of our accelerators for POSIT arithmetic.
In \ref{sec_eval}, we discuss the performance of matrix-multiplication in \POSIT{} arithmetic on FPGAs and GPUs in Section 
Section \ref{sec_lucho} is dedicated to evaluating the numerical error, the performance, and the power efficiency of matrix decomposition algorithms.
Section \ref{sec_dis} compares our FPGA designs with GPUs in \POSIT{} and in the standard binary32 arithmetic and discusses future work. 
Finally, we conclude our paper in Section \ref{sec_con}.


\section{Posit arithmetic} \label{sec_posit}

\begin{figure}
  \centering
  \includegraphics[width=\linewidth]{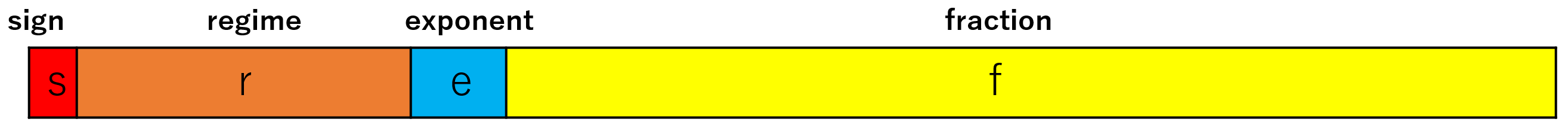}
  \caption{Bit sequence of POSIT Format}
  \label{POSITFORMAT}
\end{figure}

Figure \ref{POSITFORMAT} shows the POSIT format, where $s$ is the sign (1-bit), $r$ and $e$ are the exponent part, and $f$ is the fractional part. With the bit sequences $s, r, e$, and $f$, the value $x$ in the POSIT arithmetic is defined as follows.

\begin{equation}
x = (-1)^s \times \underline{u^{k( r )} \times 2^e} \times 1.{f},
\end{equation}
where the underlined part is the exponent part of the POSIT format, and the regime $r$ is a variable-length bit sequence.
Accordingly, the bit length of $f$ is also variable.
In the following, we call the bit length of $r$, $e$, and $f$ as $r_{\rm s}$, $e_{\rm s}$, and $f_{\rm s}$, respectively.
The factor $u$ is defined as $u = 2^{2^{e_{\rm s}}}$.
The POSIT format is identified as Posit($n_{\rm bit}$, $e_{\rm s}$) according to \cite{9139786}, where $n_{\rm bit}$ is the total bit length of a variable in the Posit format. 
Only the 32-bit POSIT format is considered in this work; hence, we use \POSIT{} and set $e_{\rm s} = 2$ and $u = 2^{2^{2}} = 16$. 

$k(r)$ is a function that decodes the  number of consecutive 1s for positive numbers or 0s for negative numbers \cite{lindstrom2018universal}.
If $x$ is positive, a sequence of consecutive 1s is terminated by ``0'' so one can use a priority encoder in hardware logic to implement the function $k(r)$.
If $x$ is negative, we use the priority encoder to count consecutive 0s terminated by ``1''.
Therefore, the bit length allocated for the fraction $f_{\rm s} = 32 - k(r) - e_{\rm s} - 2$.
The last integer ``2'' accounts for the sign and the terminating bit in the regime part.
Unlike conventional FP formats such as IEEE binary32, the maximum rounding error after each arithmetic operation is not a constant.
Depending on the magnitude of a number $|x|$, the rounding error of \POSIT{} $\epsilon_{\rm posit}$ can be smaller or larger than the machine epsilon for the binary32 format $\epsilon_{\rm binary32} \sim 6 \times 10^{-8}$.
The range of values that favor \POSIT{} is $10^{-3} < |x| < 10^{3}$. 
Otherwise,  $\epsilon_{\rm posit}$ is larger than $\epsilon_{\rm binary32}$.
This range is the golden zone for \POSITA{} \cite{9139786}.

The implementation of an FP arithmetic operation differs significantly between \POSIT{} and binary32, as \POSIT{} requires pre-processing (decoding) and post-processing (encoding) before and after an arithmetic operation.   
The exponent part of the POSIT format is first decoded to determine $f_{\rm s}$ and extract the fraction $f$ as the pre-processing step.
Then, we convert the POSIT number format into an internal FP format from their bit sequences before performing an arithmetic operation.
After the operation with the internal FP format, the size of the exponent of the internal FP format is encoded in the regime, and the fraction is rounded using the new $f_{\rm s}$ as the post-processing step.
These pre-processing and post-processing operations require loops in software implementations, with the number of iterations depending on the bit length of the exponent part.
Consequently, the number of instructions required for pre-processing and post-processing varies depending on the magnitude of a value in \POSIT{}.

\section{Details of Accelerator of Linear Algebra in \POSIT} \label{sec_acc}
We evaluate FPGAs and GPUs as \POSITA{} accelerators for linear algebra operations.
Specifically, we evaluate matrix multiplication and two decomposition algorithms for dense matrices: Cholesky and LU decompositions. 
The two decomposition algorithms are implemented in LAPACK (Linear Algebra PACKage).  
Regarding the LU decomposition in binary64, {\it dgetrf} routine in LAPACK adopts the iterative algorithm \cite{doi:10.1137/S0895479896297744} that depends on the Level 3 BLAS (Basic Linear Algebra Subroutines). 
For the Cholesky decomposition in binary64, {\it dpotrf} routine in LAPACK adopts the same iterative algorithm.
To evaluate these routines, we implement the same interface as the Level 3 BLAS routine called General Matrix Multiplication (GEMM). 
The GEMM routine performs the matrix multiplication for matrices $A$ and $B$ as follows:

\begin{equation}
C = \alpha ~op(A) ~op(B) + \beta C,
\label{GEMM_DEF}
\end{equation}
where $op(X)$ is one of $op(X) = X$ or $op(X) = X^T$, $\alpha$ and $\beta$ are scalar parameters.
For instance, the GEMM routine in binary64 in BLAS is called {\it dgemm}.

This work extends MPLAPACK~\cite{mplapack} to implement BLAS and LAPACK routines in \POSITA{}.
MPLAPACK is a re-implementation of all BLAS and LAPACK routines in C++ to support multi-precision FP operations such as the double-double (DD)~\cite{Dekker_1971, Kunuth_1998} arithmetic, and the Multiple Precision Floating-Point Reliable (MPFR)~\cite{mpfr} library.
In MPLAPACK, all BLAS and LAPACK routines are replaced with the prefix $R$, a common prefix of all routines regardless of arithmetic precision.
For instance, {\it dgemm}, {\it dgetrf}, and {\it dpotrf} are called \Rgemm{}, \Rgetrf{}, and \Rpotrf{} in MPLAPACK. 
We port only necessary routines for \Rgemm{}, \Rgetrf{}, and \Rpotrf{} in \POSIT{} arithmetic by modifying a part of MPLAPACK in our evaluation. 

\subsection{FPGA implementation}
In a hardware implementation, the logic for the pre- and post-processing is usually designed as a combinational circuit \cite{9180771,8615707,jaiswal2019pacogen}. 
Therefore, both processing is deterministic, so in principle, the hardware \POSIT{} unit has no performance penalty with arguments $|x| \gg 10^{3}$ or $|x| \ll 10^{-3}$.
We use Flo-Posit by Murillo \etal \cite{9180771} for our hardware evaluation. 
This open source project, available at \cite{FloPosit}, contains two types of \POSIT{} designs, one of which uses an internal FP format similar to the IEEE FP format, e.g., the fraction is treated as a combination of sign and unsigned magnitude as in \cite{8615707, jaiswal2019pacogen}.
Another design uses the two's complement integer as the fraction \cite{gustafson2017beating}. 
In \cite{murillo2022comparing}, they reported that the latter approach requires fewer hardware resources than the former. 

In this work, we optimize the add and multiply units with both approaches generated by Flo-Posit.
We optimize these units by inserting pipeline stages to maximize the throughput of the operations. 
We combine the pipelined add and multiply units with \fblas{}, which includes a systolic array \cite{KungSystolic} design for GEMM operations for our applications of linear algebra operations. 
Our systolic designs compute Eq.(\ref{GEMM_DEF}) without transpose operations.
To support the GEMM interface, we transpose input matrices on a host CPU before sending them to the FPGA if necessary. 

\subsection{GPU implementation}
To evaluate the performance of \POSITA{} on GPUs, we port the add and multiply routines from SoftPosit for GPU kernels in CUDA \cite{cuda} and OpenCL \cite{opencl}.
The CUDA and OpenCL kernels are identical except for differences in the kernel API.
Although modern datacenter GPUs, such as NVIDIA A100 or H100, have evolved to support faster FP performance in binary64 for traditional HPC applications, recent GPUs also target NNs with dedicated units for matrix multiplication with the arithmetic in IEEE binary16, bfloat16, and integer.
Our port of the SoftPosit routines relies only on 32-/64-bit integer instructions. 
As we will see later, this implementation is suitable for consumer GPUs, such as GeForce RTX3090, RTX4090, and RX7900XTX, which have limited support for binary64 arithmetic.
We implement optimized GPU kernels for matrix multiplication based on the add and multiply kernel functions in \POSIT{}.
As shown in Eq.(\ref{GEMM_DEF}), there are four possible combinations of the transpose operation on matrices $A$ and $B$; we have four GPU kernels for each case. 
We optimize our GPU kernels to use the shared memory of a GPU for the blocking matrix multiplication algorithm.

\section{Evaluation of \POSITA{} on FPGAs and GPUs} \label{sec_eval}

\subsection{Evaluation of GEMM on FPGAs} \label{FPGA_IMP}
For the evaluation of FPGAs, we use the Terasic DE10a-Net Agilex board (hereafter Agilex). 
The Terasic DE10a-Net has an Agilex FPGA chip (AGFB014R24B2E2Vxs) and four 8GB DDR4-2666 DIMM modules. 
We use Intel's OpenCL-based High-Level Synthesis (HLS) to design our \POSIT{} accelerator for the GEMM operations.
The FPGA design software is Quartus 21.2.0 Build 72. 


The peak performance of GEMM depends on the number of processing elements (PEs) in a systolic array and the clock frequency of the design.
When we have $n_{\rm PE}$ PEs in our design running at $f$ MHz (Fmax), the peak performance $F_{\rm peak}$ in giga FP operations per second (Gflops)
is given by Eq.(\ref{peak}).
\begin{equation}
F_{\rm peak} = 2\ n_{\rm PE}\ f \times 10^{-3}.
\label{peak}
\end{equation}
Although we have explored different configurations of $n_{\rm PE}$, we present only the largest configuration where $n_{\rm PE} = 16 \times 16 = 256$.
For the logic synthesis and the place-and-route operation, we ran at least five times with a different seed for Quartus and selected the best design with the highest Fmax.

In Table~\ref{Agilex}, we present the two best designs for Agilex: (1) the design using the sign-and-magnitude format as the internal FP format as \POSIT{}$_{\rm SM}$ in the first column and (2) the design using the two's-complement format as the internal FP format as \POSIT{}$_{\rm TC}$ in the second column. 
In both designs, the logic cell is the bottleneck.
\POSIT{}$_{\rm TC}$ is more resource efficient than \POSIT{}$_{\rm SM}$ regarding the logic cell, while the Fmax of \POSIT{}$_{\rm TC}$ is about the same as that of \POSIT{}$_{\rm SM}$.
This result is consistent with \cite{murillo2022comparing}. 
Since the Fmax of the two designs is about the same, $F_{\rm peak}$ is also comparable. 
The last row shows the total on-chip power dissipation obtained from the {\it quartus\_pow} command.
We set the toggle rate at 25\% to estimate the power dissipation.
Considering that the power dissipation of the on-chip memory (four 8GB DIMM modules) is estimated to be 20W, the total power consumption of our design is at most 60 watts.
In the following, we present the evaluation of \POSIT{}$_{\rm TC}$, since we obtain the same numerical results from \POSIT{}$_{\rm TC}$ and \POSIT{}$_{\rm SM}$ designs and observe no practical difference in the performance. 

Table~\ref{Agilex} also compares the same systolic array with 256 PEs equipped with the binary32 FP unit.
The third and fourth columns show the hardware resources of the binary32 arithmetic.
Since it is possible to configure each DSP of the Agilex FPGA as a multiply-add unit in the binary32 arithmetic, we implement two versions of the systolic array as binary32$_{\rm Hard}$ and binary32$_{\rm Soft}$.
For binary32$_{\rm Hard}$, we use the hardware multiply-add unit of the DSP.
For binary32$_{\rm Soft}$, we use add and multiply units generated by FloPoCo \cite{DinechinPasca2011-DaT}. 
Since binary32$_{\rm Hard}$ uses the FP unit in the DSP, it requires the least resources regarding the logic cells. 
Comparing binary32$_{\rm Soft}$ and \POSIT{}$_{\rm TC}$, the latter requires more logic cells mainly due to the pre- and the post-processing.
The last row shows the total on-chip power dissipation with the toggle rate at 25\%.
The impact of the more complex logic of \POSIT{}$_{\rm TC}$ on power dissipation is modest.

\begin{figure}[h]
\centerline{\includegraphics[width=\linewidth]{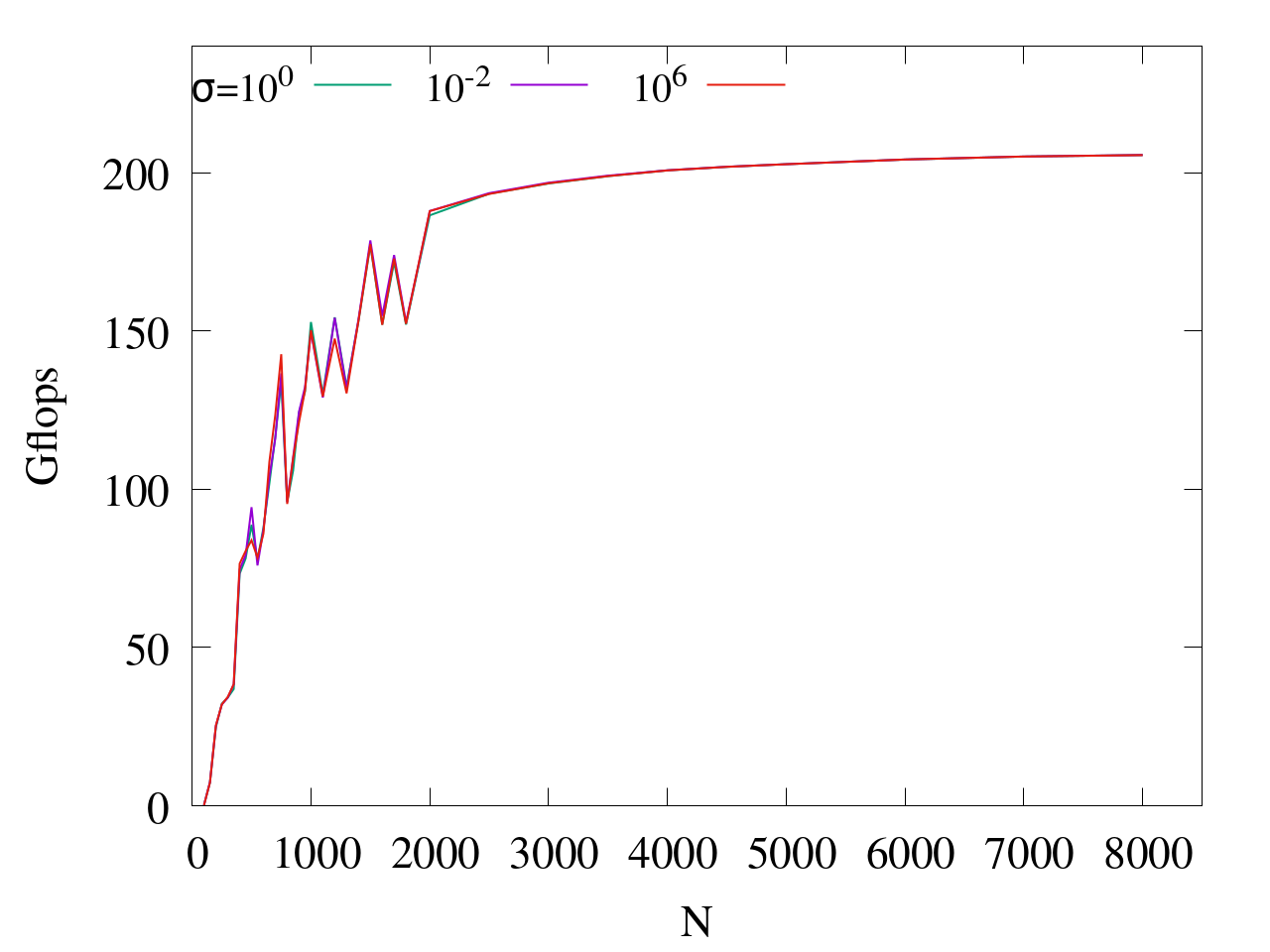}}
\caption{Performance of GEMM on Agilex with $\sigma = 10^{0}, 10^{-2},$ and $10^{6}$ for generating the square matrices.} 
\label{gemm_fpga}
\end{figure}

We measure the execution time of the \POSIT{}$_{\rm TC}$ design on Agilex by scaling $N$, where we compute the matrix multiplication for $N\times N$ square matrices $A$ and $B$. 
In this case, the number of operations of GEMM is $\sim 2 N^3$. 
Figure \ref{gemm_fpga} shows the performance of GEMM as a function of $N$ without transpose operations.
The input matrices are initialized with random numbers drawn from normal distributions with mean $\mu = 0.0$ and standard deviation $\sigma = 10^{-2}$, $10^{0}$ and $10^{6}$.
The three lines are mostly overlapped, so the performance does not depend on $\sigma$ as expected for Agilex.


\begin{table*}[t]
\begin{center}
\caption{Synthesis results of the GEMM designs in \POSIT{} and binary32 on Agilex FPGA} \label{Agilex}
\small
\begin{tabular}{ccccc}
\toprule
  & \POSIT{}$_{\rm SM}$ & \POSIT{}$_{\rm TC}$ & binary32$_{\rm Hard}$ & binary32$_{\rm Soft}$             \\
\midrule
Logic cells& 433,836 ( 89 \% )    & 337,111 ( 69 \% )    & 141,930 ( 29 \% )    & 234,697 ( 48 \% )     \\
DSP blocks & 589 ( 13 \% )        & 589 ( 13 \% )        & 317 ( 7 \% )         & 589 (13 \% )          \\
Memory bits& 15,907,652 ( 11 \% ) & 15,907,652 ( 10 \% ) & 15,890,756  ( 11 \% ) & 15,907,652 ( 11 \% ) \\
RAM blocks &  1,364 ( 19 \% )     & 1,364 ( 19 \% )      & 1,362 ( 19 \% )     & 1,364 ( 19 \% )        \\
Fmax (MHz) &  432.71              &  429.92              & 505.05              & 461.46                 \\
$F_{\rm peak}$ (Gflops) & 221.5   &  220.1               & 285.6               &  236.3                 \\
Power (watts) & 42.1  & 38.7 & 31.6  & 36.0  \\
\bottomrule
\end{tabular}
\end{center}
\end{table*}

\subsection{Evaluation and Analysis of \POSITA{} on GPUs}
As mentioned before, the software implementation of the \POSITA{} requires a loop for the function $k(r)$.
This loop counts consecutive 1s or 0s by testing each bit sequentially.
Therefore, the number of instructions needed for the function $k(r)$ depends on the absolute values of the two arguments of the operation.
This aspect is critical to the \POSITA{} in GPU implementation.
Furthermore, a GPU kernel is executed in multiple threads in lockstep, even if branch instructions are included.
Therefore, to examine the impact of the magnitude of arguments in \POSIT{} kernels, we measure the computation time of the kernels by changing the magnitude of the arguments. 

Consider a range $[a,b) = \{x | a \le x < b \}$, within which we generate random values for the arguments.
The ranges of five patterns, $I_0,I_1,I_2,I_3,I_4,$ are shown in Table~\ref{tabrandom}.
For example, $f_{\rm s}$ of a value in the range $I_0$ is $27$, the largest $f_{\rm s}$ of all ranges. 
Consequently, the corresponding $\epsilon_{\rm posit}$ is $= 2^{-27} \sim 7.5 \times 10^{-9}$.
On the other hand, when the magnitude of a value is small ($I_1$) and large ($I_2$), the $f_{\rm s}$ ranges from 0 to 3 bits, and $\epsilon_{\rm posit} = 2^{-3}$ at most.

\begin{table*}[t]
\begin{center}
\caption{The elapsed time (in nano-second) of GPU kernels for a given range of input arguments}
\label{tabrandom}
\begin{tabular}{cllcccc}
\toprule
& $a$ & $b$ & Add & Mul & Div & Sqrt \\
\midrule
$I_0$ & 1.0        & 2.0        & 101 & 101 & 173 & 96  \\
$I_1$ & $10^{-38}$ & $10^{-30}$ & 215 & 209 & 301 & 143 \\
$I_2$ & $10^{30}$  & $10^{38}$  & 210 & 209 & 309 & 148 \\
$I_3$ & $10^{-15}$ & $10^{-14}$ & 148 & 141 & 233 & 136 \\
$I_4$ & $10^{14} $ & $10^{15}$  & 145 & 141 & 230 & 136 \\
\bottomrule
\end{tabular}
\end{center}
\end{table*}

\begin{table*}[t]
\begin{center}
\caption{Profiling results of Add kernel for a given range of input arguments}
\label{tabprof}
\begin{tabular}{cllcccc||ccc}
\toprule
& $n_{\rm inst}$ & $n_{\rm cont}$ & $f_{\rm branch}$ \\
\midrule
$I_0$ &  81  & 26 & 94.74 \% \\
$I_1$ &  283 & 73 & 93.04 \% \\
$I_2$ &  237 & 76 & 93.95 \% \\
$I_3$ &  175 & 46 & 91.04 \% \\
$I_4$ &  150 & 46 & 91.83 \% \\
\bottomrule
\end{tabular}
\end{center}
\end{table*}

For the ranges shown in Table~\ref{tabrandom}, we measure the execution time of \POSIT{} kernels in nanoseconds for add (Add), multiply (Mul), divide (Div), and square root (Sqrt) operations on NVIDIA V100 GPU.
The Add, Mul, and Div kernels perform the corresponding arithmetic operation on input arrays $p[S]$ and $q[S]$, where $S$ represents the size of the arrays. 
For instance, the addition kernel performs $ r[i] = p[i] + q[i]$ in \POSITA{} for $i = 0\ ...\ S-1$.
We set $S$ to a sufficiently large value, such as $10^5$, to measure the effective kernel time per CUDA core.
We use the array $p[S]$ as the input for the Sqrt kernel.

The execution time per CUDA core is shown in Table~\ref{tabrandom} in the columns Add, Mul, Div, and Sqrt.
Each column shows the execution time per CUDA core with the arrays $p$ and $q$ initialized by values randomly drawn from the given ranges $I_0$ to $I_4$.
By examining Table~\ref{tabrandom}, we observe that the performance is the worst for the ranges $I_1$ and $I_2$, where $f_{\rm s}$ is the smallest, and the performance is the best for the range $I_0$.  
Therefore, when using these \POSIT{} kernels on GPUs, making the equations dimensionless proves effective from both a computational error and performance perspective. 


For the same ranges as Table~\ref{tabrandom}, Table~\ref{tabprof} shows the results of instruction profiling when the Add kernel is executed.
We use the command {\it nvprof} to profile the GPU kernels. 
The column $n_{\rm inst}$ shows the number of all instructions executed by the kernel per operation.
The execution time in the column Add in Table~\ref{tabrandom} is proportional to $n_{inst}$. 
The $n_{\rm cont}$ column shows the number of executed control instructions, such as branch instructions.
The larger $n_{\rm cont}$ means the kernel executes more number of control instructions in the kernel.
With the ranges of $I_1$ and $I_2$, where the argument range is wide, $n_{\rm cont}$ is the largest, as expected. 
The column $f_{\rm branch}$ shows the branch efficiency, the percentage of threads in a GPU warp without branch divergence. 
If $f_{\rm branch} = 100 \%$, all threads in the warp took the same path in branching instructions.
The smaller $f_{\rm branch}$, the longer the execution time of the GPU kernel is.
In the case of the Add kernel, $f_{\rm branch}$ is the worst in the argument ranges $I_3$ and $I_4$.
The instruction profiling results for other kernels show the same trend as that for the Add kernel.
To summarize, the performance of \POSIT{} arithmetic on GPUs is not constant but depends on the absolute value of input arguments.

\begin{table*}[t]
\begin{center}
\caption{Specification of GPUs} \label{GPUs}
\begin{tabular}{lccccc}
\toprule
                       & V100 & H100 & RTX3090 & RTX4090 & RX7900 \\ 
\midrule
Process node(nm)       & 12 & 4 & 8 & 5 & 5 \\
Number of cores            & 5120 & 14592 & 10496 & 16384  & 6144 \\
Clock(MHz)             & 1245 & 1065 & 1400 & 2235  & 1855 \\
Memory(GB)             & 32 & 80 & 24 & 24 &24 \\
Tops(integer)          & 6.37 & 15.5 & 14.7 & 36.6 & 22.8 \\
Tflops(binary32)       & 14  & 51 & 36    & 83  & 61  \\
Tflops(binary64)       & 7.1 & 25 & 0.56  & 1.3 & 1.9  \\
$P_{\rm limit}$(watts) & 250 & 360 & 350 & 450 & 339 \\
\bottomrule\end{tabular}
\end{center}
\end{table*}

\subsection{Evaluation of GEMM on GPUs}
This section presents the performance of GEMM kernels on GPUs presented in Table~\ref{GPUs}.
We use NVIDIA V100 PCIe (GV100; Compute Capability 7.0), NVIDIA H100 Tensor Core GPU with PCIe (GH100; Compute Capability 9.0), NVIDIA GeForce RTX3090 (GA102; Compute Capability 8.6), NVIDIA GeForce RTX4090 (AD102; Compute Capability 8.9), and AMD Radeon RX7900XTX (Navi31; hereafter RX7900). 
Table~\ref{GPUs} shows the specifications of these GPUs.
The number of cores is the number of CUDA cores for the NVIDIA GPUs and the number of Stream Processors for the AMD GPU.
The ``Clock'' is the base clock speed for each GPU.
In the ``Tops'' row, we estimate the number of 32-bit integer operations in tera operations per second (Tops) for each GPU, assuming that each core performs one (two for the AMD GPU) 32-bit integer operations at the base clock.
$P_{\rm limit}$ is the default power limit in watts.

\begin{figure}[h]
\centerline{\includegraphics[width=\linewidth]{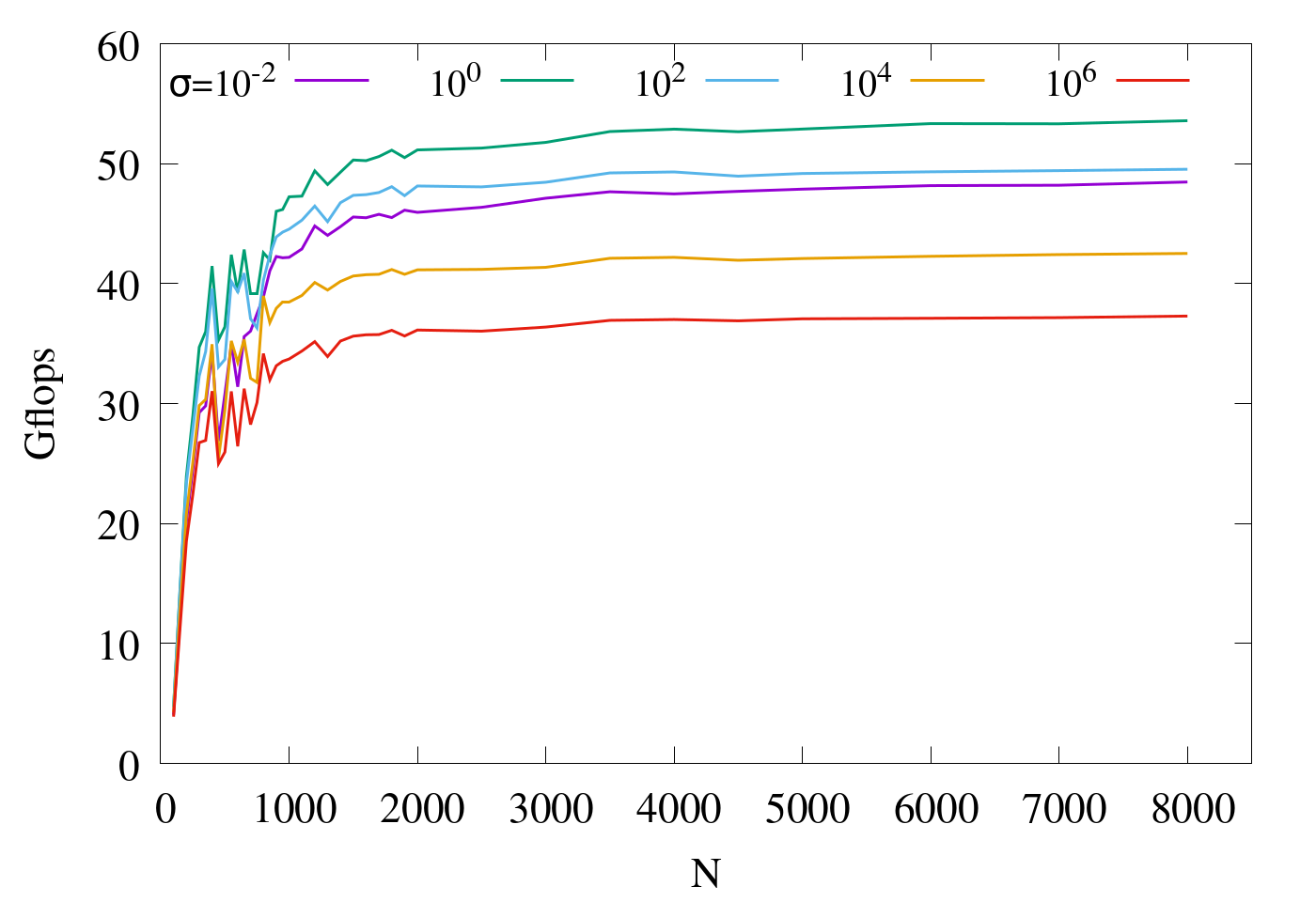}}
\caption{Performance of GEMM on V100 GPU with different $\sigma$ for generating the square matrices.}
\label{gemm_v100}
\end{figure}

We execute the same OpenCL GEMM kernels on these GPUs and measure the execution time in the same way as the evaluation of FPGA designs. 
Figure \ref{gemm_v100} shows the performance as a function of $N$ without transpose operations for V100.
The input matrices are initialized with random numbers drawn from normal distributions with mean $\mu = 0.0$ and standard deviation $\sigma = 10^{-2}, 10^{0}, 10^{2}, 10^{4}$, and $10^{6}$.
As expected, the performance of GEMM depends on the magnitude of the elements of the input matrices.
For $\sigma = 10^{0}$, the highest performance is $\sim 55$ for V100.
The performance degradation is significant for both smaller and larger $\sigma$.
For example, the performance with $\sigma = 10^{6}$ is only $\sim 37$ Gflops.

\begin{figure}[h]
\centerline{\includegraphics[width=\linewidth]{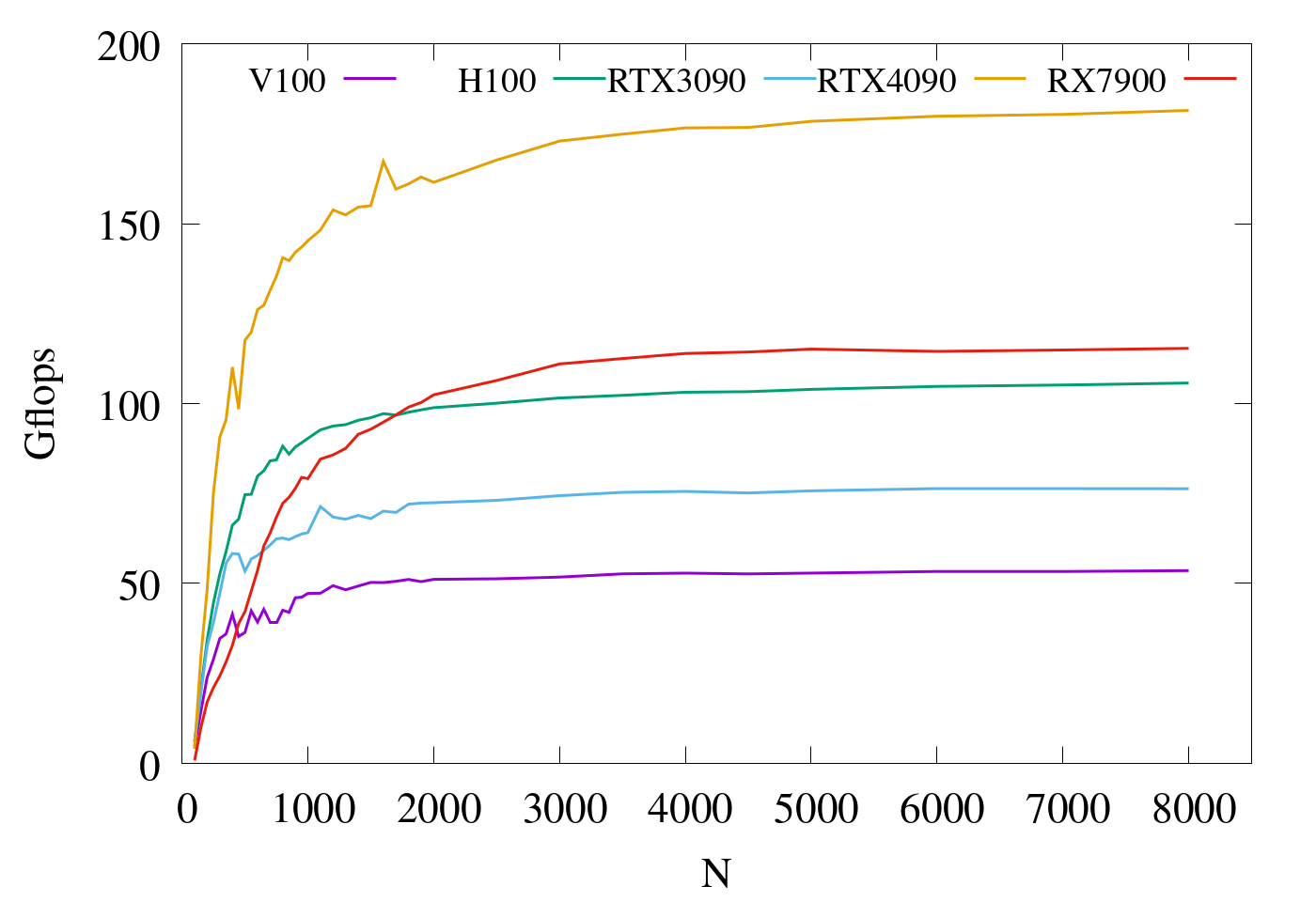}}
\caption{Performance of GEMM on five GPUs with $\sigma = 10^{0}$.}
\label{gemm_gpu_all}
\end{figure}

Figure \ref{gemm_gpu_all} shows the performance of GEMM on five GPUs.
Note that $P_{\rm limit}$ of the GPUs are different, as shown in Table~\ref{GPUs}.
Comparing these GPUs, RTX4090 and RX7900 are faster than the datacenter GPUs. 

\begin{figure}[h]
 \begin{minipage}[b]{\linewidth}
 \centerline{\includegraphics[width=\linewidth]{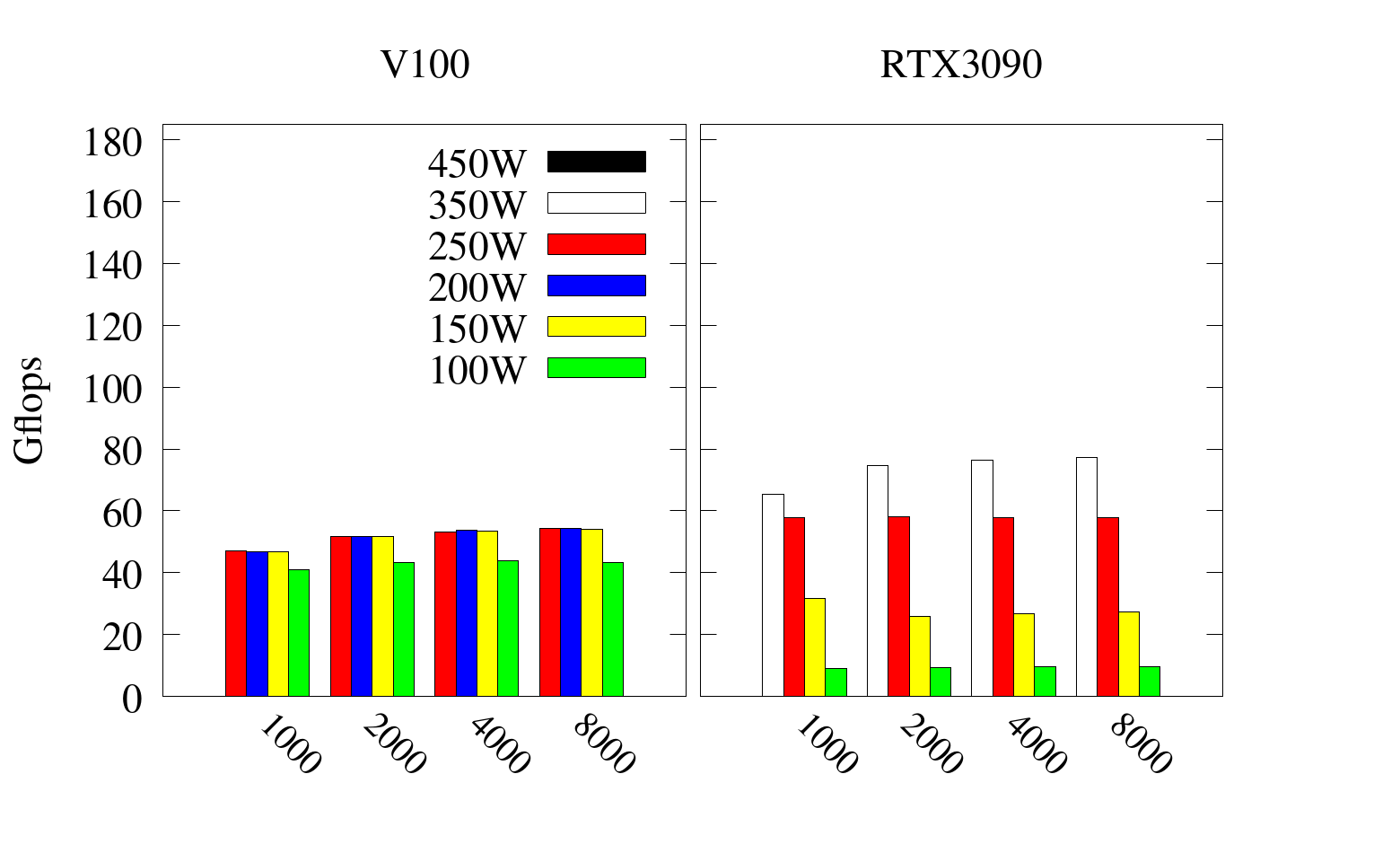}}
 \end{minipage}\\
 \begin{minipage}[b]{\linewidth}
 \centerline{\includegraphics[width=\linewidth]{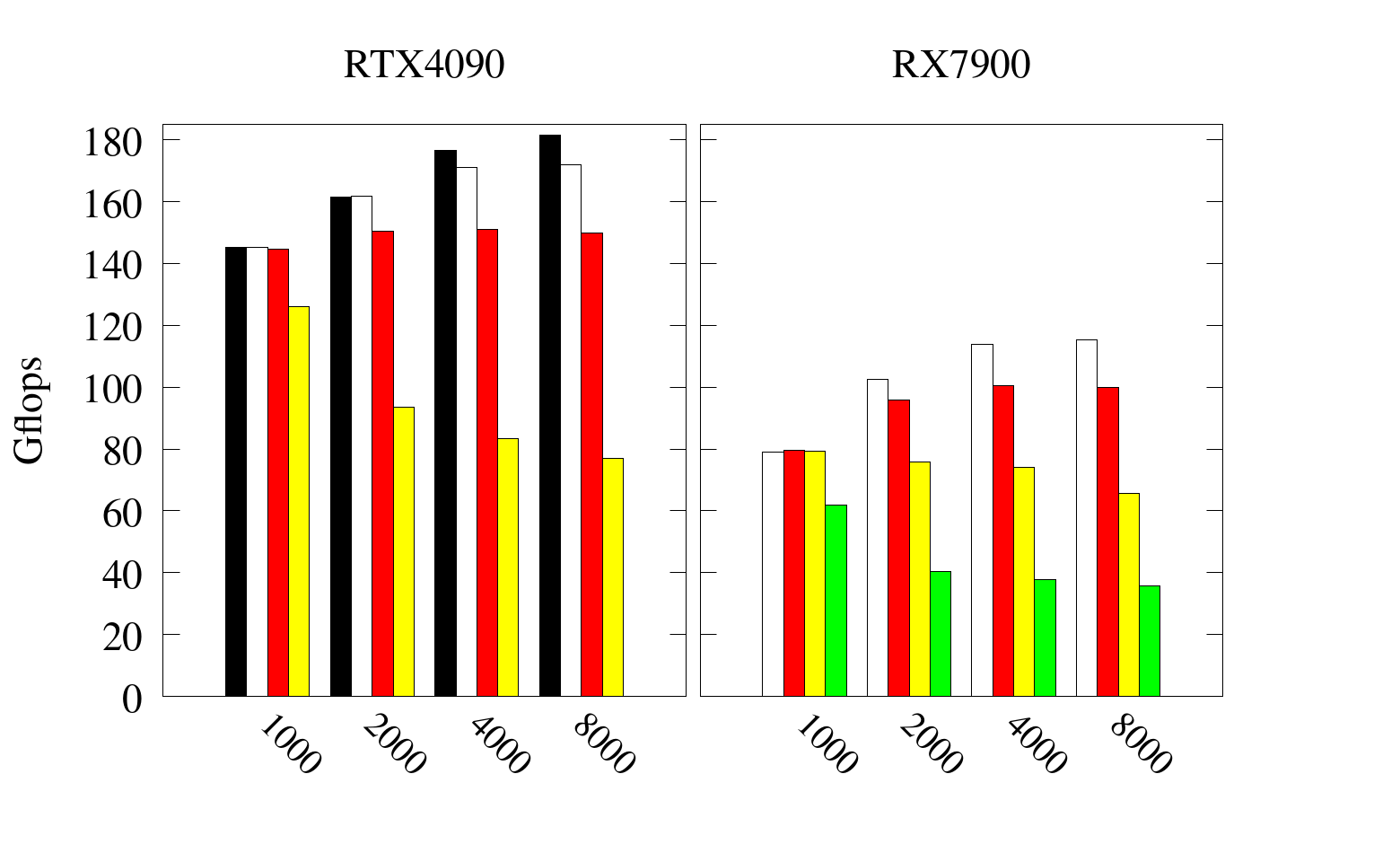}}
 \end{minipage}
\caption{Performance of GEMM on V100, RTX3090, RTX4090 and RX7900 with different $P_{\rm limit}$. The performance measured with $P_{\rm limit}$ = 450, 350, 250, 150, and 100 watts are shown as black, white, red, blue, yellow, and green bars, respectively. We set $\sigma = 10^{0}$.
  }
\label{gemm_gpu_pl}
\end{figure}

For $\sigma = 10^{0}$, the highest performance for V100 and RTX4090 is $\sim 55$ and $181$ Gflops, respectively.
The performance difference between the two GPUs is about $125$ Gflops.
To investigate how the performance of GEMM depends on $P_{\rm limit}$, we vary $P_{\rm limit}$ and measure the execution time of the GEMM kernels on four GPUs except H100. 
We initialize the input matrices in these evaluations with the normal distribution $\mu = 0.0$ and $\sigma = 10^{0}$.
Figure \ref{gemm_gpu_pl} shows the performance as a function of $N$ for V100, RTX3090, RTX4090, and RX7900. 
For V100, the performance is almost the same for $P_{\rm limit}$ = 250, 200, and 150 watts.
At $P_{\rm limit} = 100$ watts, the performance drops to $\sim$ 40 Gflops at $N = 8000$.
However, the performance of RTX3090, RT7900, and RTX4090 is highly dependent on $P_{\rm limit}$.
With the same $P_{\rm limit} = 250$ watts, the performance of the three GPUs is $\sim 58$, $100$, and $150$ Gflops, respectively at $N = 8000$.
With the lower power limit at $P_{\rm limit} = 150$ watts,  the performance of the three GPUs is $\sim 27$, $66$, and $77$ Gflops, respectively.
At $P_{\rm limit} = 250$ watts, the gap in the performance difference between V100 and RTX4090 is $\sim 100$ Gflops. 
At $P_{\rm limit} = 100$ watts, the gap in the performance difference between V100 and RTX4090 is reduced to $20$ Gflops.
We suspect that the performance difference between the datacenter GPUs and the consumer GPUs is due to the difference in their computational capabilities and detailed differences in the GPU power limit control mechanism.


\subsection{Comparison of GEMM performance on FPGAs and GPUs}
Comparing the results with Agilex and five GPUs, the performance of Agilex is higher at a large $N$.
At $N = 8000$, the performance of Agilex is 202.7 Gflops, while that of RTX4090 is 181.4 Gflops. 
However, at $N < 3000$, the performance of Agilex is a less steep function of $N$ than that of the GPUs.
The performance of all GPUs peaks around $N \sim 2000$, while the full potential of Agilex is ineffective at $N < 3000$.
While Agilex has a PCI-Express Gen3x16 interface, all GPUs have a PCI-Express Gen4x16 interface.
Accordingly, the data transfer from the host CPU to the FPGA is the severe bottleneck for small matrices.

\begin{figure}[h]
\centerline{\includegraphics[width=\linewidth]{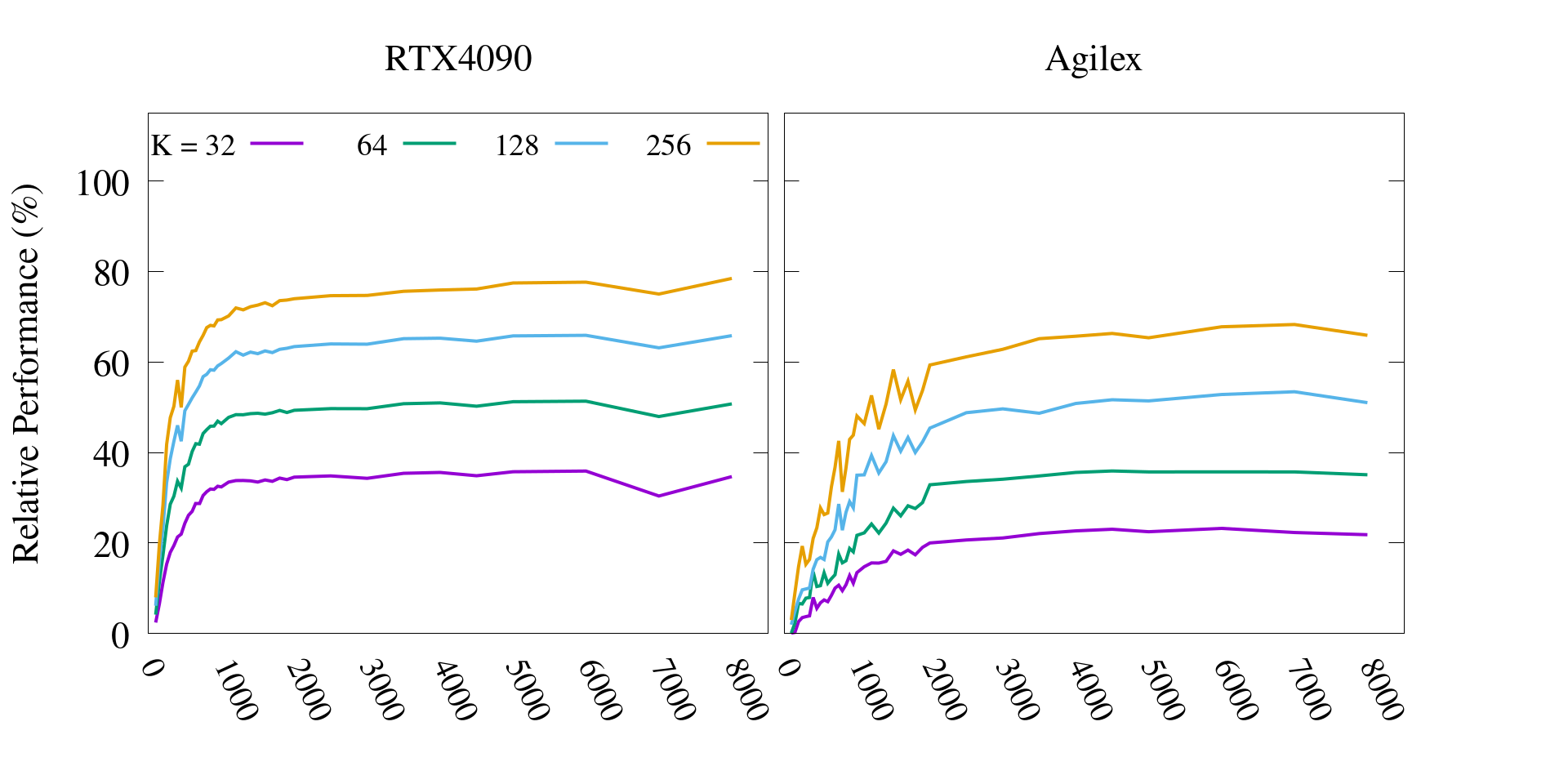}}
\caption{Performance of GEMM relative to $F_{\rm peak}$ of RTX4090 and Agilex FPGA on non-square matrices $AB$, where the matrix $A$ is an $N$ by $K$ matrix and the matrix $B$ is a $K$ by $N$ matrix.}
\label{gemm_trailing}
\end{figure}

Another weakness of GEMM on FPGA designs is that the performance of GEMM on non-square matrices is lower than that on square matrices, as shown in Figure \ref{gemm_trailing} and reported by \cite{2023arXiv230604087K}.
We evaluate the performance of GEMM on non-square matrices $AB$, where the matrix $A$ is an $N$ by $K$ matrix and the matrix $B$ is a $K$ by $N$ matrix.
Here, the multiplication result $AB$ is an $N$ by $N$ matrix, and the number of operations of GEMM is $\sim 2 N^2 K$. 
These GEMM operations, called the trailing matrix update, are used in a blocked matrix decomposition algorithm such as the LU decomposition.
The trailing matrix update is less computationally intensive when compared to GEMM operations on square matrices.

Figure \ref{gemm_trailing} shows the performance relative to $F_{\rm peak}$ of RTX4090 and Agilex FPGA.
For this evaluation, we set $\sigma = 10^{0}$.
For RTX4090, we set $F_{\rm peak} = 181.5$ Gflops, which is the performance of GEMM for square matrices at $N = 8000$.
We find that the performance of the trailing matrix update on Agilex is only 20\% at $K = 32$.
On RTX4090 and other GPUs, the relative performance of the trailing matrix update is better than that of Agilex for the evaluated values of $K$.

In the current systolic architecture, $16 \times 16$ PEs are arranged as a mesh, with each PE performing the multiply operation followed by the add operation.
With our optimized \POSIT{} units, the total number of clock cycles of the two operations at each PE is 11 cycles.
Accordingly, the number of clock cycles along a row or a column of PEs is at least 176.
If the dimension of the matrices $A$ and $B$ is small or $K < N$ as the trailing matrix update, the latency of the PE pipeline makes the utilization of the PEs low.
We do the same evaluation on a smaller systolic array of $8 \times 8$ PEs.
With $K = 32$ and $N > 2000$, the performance of the array is $\sim 27$ Gflops, which is more than 50\% of its $F_{\rm peak}$.
With $K = 256$ and $N > 2000$, the performance of the array relative to its $F_{\rm peak}$ is close to 100\%, so that the small array is much more effective for the trailing matrix update.

\section{Evaluation of Matrix Decomposition on FPGAs and GPUs}\label{sec_lucho}
\subsection{Evaluation of Numerical Error of Matrix Decomposition of \POSITA{}}
To evaluate the numerical error of \POSITA{}, we use the same method as in \cite{GHYSELS2014224,9139786}.
We use the factorized matrix to solve a linear equation $A\vec{x} = \vec{b}$ using \Rpotrs{} and \Rgetrs{} for the Cholesky and LU decompositions, respectively.
Using binary64 arithmetic, we set the true solution of the linear equation as $\vec{x}_{\rm sol}$ with all components being $1/\sqrt{N}$ and compute $\vec{b}_{\rm binary64} = A\vec{x}_{\rm sol}$ as the right-hand side of the linear equations. 
With the obtained numerical solution $\vec{x}_{\rm \Rpotrs}$ for the Cholesky decompositions, we evaluate the numerical error as the relative backward error as
\begin{equation}
e_{\rm \Rpotrs} = \frac{|\vec{b}_{\rm binary64} - A \vec{x}_{\rm \Rpotrs}|}{|\vec{b}_{\rm binary64}|}.
\end{equation}
We also solve the same linear equation in binary32 arithmetic by calling {\it Spotrs} from the LAPACK library.
Let the relative backward error obtained by binary32 arithmetic $e_{\rm {\it Spotrs}}$, and we quantify the relative advantage of \POSITA{} over binary32 arithmetic with the ratio between the two errors as
\begin{equation}
{\rm log_{10}} \frac{e_{\rm Spotrs}}{e_{\rm Rpotrs}}.
\end{equation}
This value is positive if \POSITA{} is more accurate than binary32 arithmetic in relative digits.
Finally, we do the same error analysis for the LU decomposition by solving the linear equation with {\it Sgetrs} in binary32 arithmetic.
The results presented here are evaluated using V100 because we have the same numerical results on GPUs and FPGAs.

\begin{figure}[h]
  \centerline{\includegraphics[width=\linewidth]{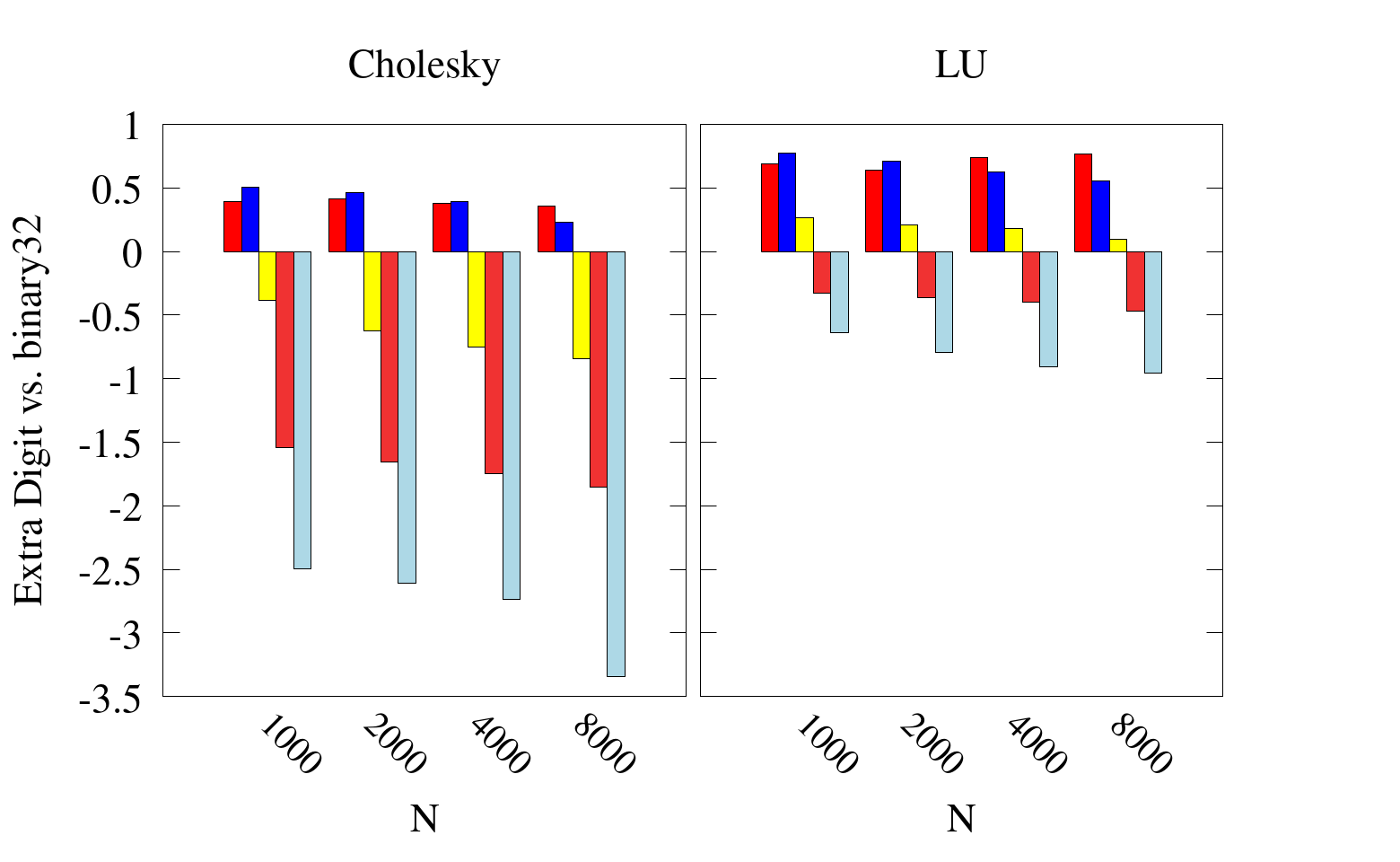}}
  \caption{The relative advantage of \POSIT{} over binary32 in the number of digits for the Cholesky and LU decompositions. For each bar from left to right, the values are evaluated for the input matrix generated with $\sigma = 10^{-2}, 10^{0}, 10^{2}, 10^{4}$, and $10^{6}$.}
  \label{cho_err}
\end{figure}

Figure ~\ref{cho_err} shows the ratio as a function of $N$ with $\sigma$ for the two decompositions.
In both cases, we alter the input matrix with $\sigma = 10^{-2}, 10^{0}, 10^{2}, 10^{4}$, and $10^{6}$.
With larger $\sigma$, the relative advantage of \POSITA{} is smaller.
When $\sigma >= 10^{2}$, \POSITA{} has no advantage.
The results for \Rpotrf{} are more severely affected by a large norm of the elements of the input matrix than that for \Rgetrf{}.
As reported in \cite{9139786}, where the authors have evaluated the numerical stability of \POSITA{} for sparse matrices, 
scaling $A$ and $b$ by a factor that makes the absolute values of the elements of $A$ and $b$ as close to 1 as possible
is effective to improve the accuracy of \POSITA{}.

\subsection{Performance of Matrix Decomposition}
\begin{figure}[h]
  \centerline{\includegraphics[width=\linewidth]{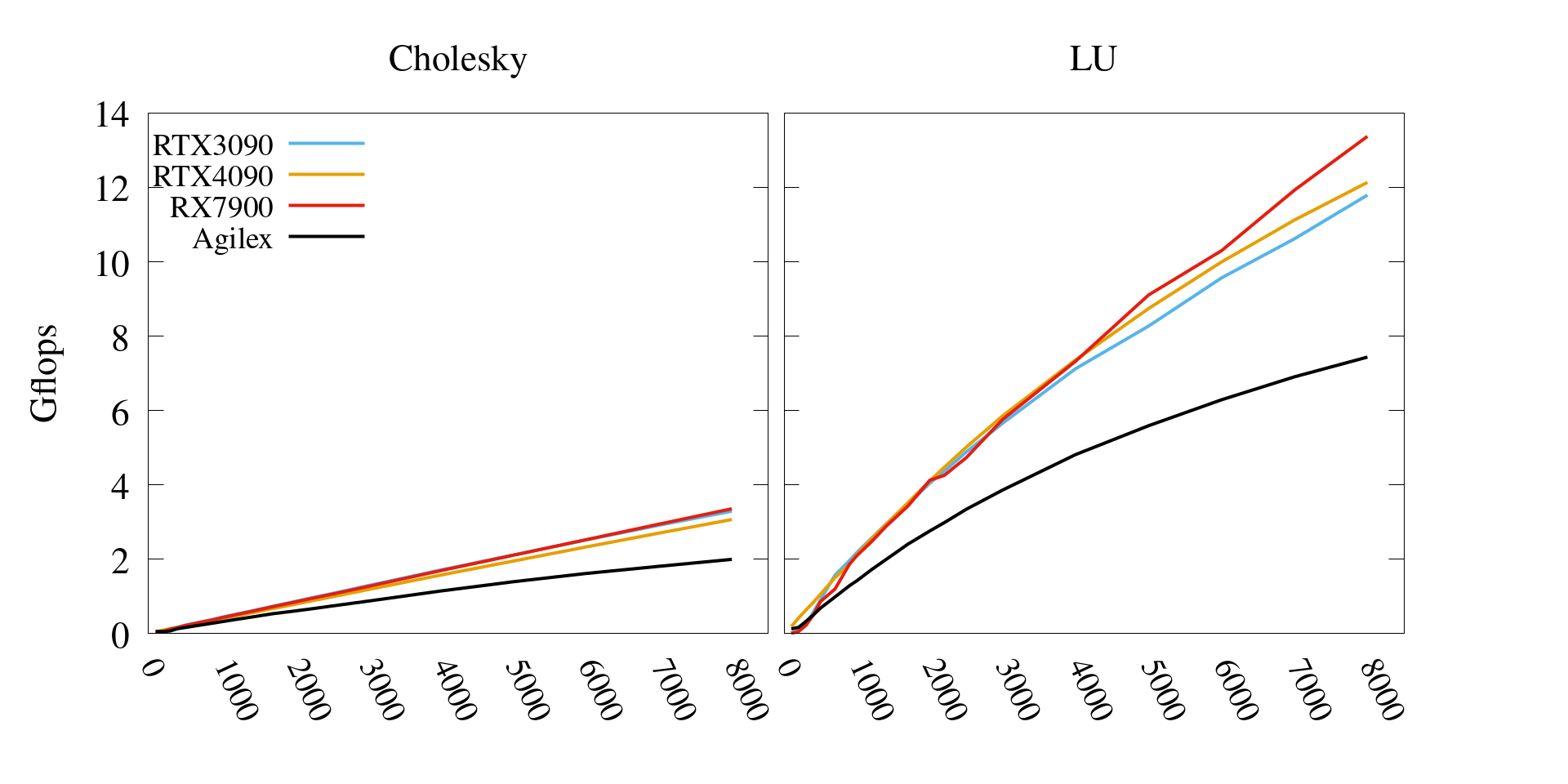}}
  \caption{Performance of the Cholesky and LU decompositions on three consumer GPUs and Agilex FPGA.}
  \label{cho_lu_flops}
\end{figure}

To evaluate the Cholesky and the LU decompositions in \POSITA{}, we execute \Rpotrf{} and \Rgetrf{} using RTX3090, RTX4090, RX7900, and Agilex FPGA.
Both \Rpotrf{} and \Rgetrf{} call \Rgemm{} for updating the trailing matrix. 
We run the two decomposition routines on an $N \times N$ matrix. 
Since \Rpotrf{} factorizes a Hermitian and positive definite matrix, we initialize a matrix $X$ with the normal distribution and compute a matrix $A = X^{T}X$ as the input matrix.
By measuring the execution time of the two routines, we compute the performance with the number of operations as $\frac{N^3}{3}$ and $\frac{2 N^3}{3}$ for \Rpotrf{} and \Rgetrf{}, respectively. 
In Figure~\ref{cho_lu_flops}, we present the performance of \Rpotrf{} and \Rgetrf{} in Gflops as a function of $N$.
In both cases we initialize the input matrix with $\sigma = 10^{0}$. 
Since the LU decomposition is more compute-intensive than the Cholesky decomposition, the performance of \Rgetrf{} is much better than that of \Rpotrf{} in all systems.

Table~\ref{taball} shows the elapsed time for the two decomposition algorithms with and without accelerators at $N = 8000$.
The column $n_{\rm core}$ shows the number of CPU cores used. 
Since we only accelerate \Rgemm{} routine with either GPU or FPGA, the performance of a CPU also affects the elapsed time. 
The first row shows the results obtained with Agilex as the accelerator.
The next five rows show the results obtained using GPUs as accelerators.
The seventh, eighth, and ninth rows with the asterisk show the results obtained using the consumer GPUs with $P_{\rm limit} = 150$ watts for RTX4090
and $P_{\rm limit} = 100$ watts for RX7900 and RTX3090.
These values are the lowest $P_{\rm limit}$ for each GPU.
Other rows show the results obtained using CPU only on various systems with the OpenMP parallelization of \Rgemm{} enabled.
We use Core i9-10900 as the host CPU for Agilex FPGA system, Xeon Gold 5122 for V100, Xeon Platinum 8468 for H100, Core i9-13900K for RTX4090, and Ryzen9 7950X for RTX3090 and RX7900.

Setting the lowest $P_{\rm limit}$ does not degrade the performance of RTX4090 and RX7900, but RTX3090 is about three times slower than that with the default $P_{\rm limit} = 350$ watts as shown in Table~\ref{taball}.
In fact, at $N = 8000$, the GPU utilization during the execution of the two decomposition algorithms do not peak out due to the nature of the algorithms.

The performance without accelerators is also promising when we have more $n_{\rm core}$, because the performance of \Rgemm{} scales well with the clock speed of a CPU and $n_{\rm core}$ of the CPU.
For example, both Ryzen 9 7950X and EPYC 7313P have $n_{\rm core} = 16$, but the base clock of the former CPU, 3.0 GHz, is higher than the latter.
Accordingly, Ryzen9 is the fastest among all CPUs evaluated in our evaluations.


\begin{table*}[t]
\begin{center}
\caption{Elapsed time (second) for solving the two decompositions at $N = 8000$}
\label{taball}
\begin{tabular}{lrrrc}
\toprule
              & Cholesky & LU  & $n_{\rm core}$ & Use FPGA/GPU \\
\midrule
Agilex           & 85.6   & 45.9  & 10 & $\checkmark$  \\
RX7900           & 50.9   & 25.5  & 16 & $\checkmark$  \\ 
RTX3090          & 51.9   & 28.9  & 16 & $\checkmark$  \\
RTX4090          & 55.7   & 28.1  & 24 & $\checkmark$  \\
H100             & 102.2  & 46.2  & 24 & $\checkmark$  \\
V100             & 115.1  & 56.2  & 4  & $\checkmark$  \\
RTX4090$^{*}$    & 55.5   & 28.1  & 24 & $\checkmark$  \\ 
RX7900$^{*}$     & 49.2   & 25.5  & 16 & $\checkmark$  \\ 
RTX3090$^{*}$    & 64.9    & 61.9  & 16 & $\checkmark$  \\
Ryzen9 7950X       & 144.9  & 207.4   & 16 &        \\ 
Core i9-13900K     & 150.2  & 243.8   & 24 &        \\
EPYC 7313P         & 280.0  & 443.6   & 16 &        \\
Core i9-10900      & 620.0  & 1042.2  & 10 &        \\
\bottomrule
\end{tabular}
\end{center}
\end{table*}

\subsection{Power Efficiency of the LU decomposition}
Finally, we present the evaluation of the power consumption of Agilex and GPUs as we solve the LU decomposition algorithm.
During the repeated execution of the LU decomposition algorithm, we monitor the AC power consumption of each system.
Based on the AC power consumption record, we calculate the average power consumption of each system.
The results are shown in Table~\ref{PowerE}.
Because we use different CPUs, memory types, CPU coolers, and power supply units in each system,
the four systems are not completely in the same condition. 
Also note that Agilex FPGA only works correctly on Intel CPUs due to unknown technical issues at the time of writing.

At $N = 8000$, the LU decomposition performance of Agilex is the slowest. 
The system with Agilex as the accelerator is more power efficient at $0.050$ Gflops/watts than the system with RTX3090.
However, RX7900 and RTX4090 are more power efficient than Agilex at $0.076$ and $0.058$ Gflops/watts, respectively, 
While the process node size of the Agilex chip is 10nm, manufactured by Intel, the process node size of the two GPUs is 5nm, manufactured by Taiwan Semiconductor Manufacturing Company.
The latter process technology is newer and more advanced.

\begin{table*}[t]
\begin{center}
\caption{Power Efficiency of Agilex and GPUs for the LU decomposition at $N = 8000$}
\label{PowerE}
\begin{tabular}{lcccc}
\toprule
                 & Agilex & RTX3090 & RTX4090 & RX7900 \\
\midrule
Process node (nm) of FPGA/GPU & 10 & 8 & 5 & 5 \\
Board Memory (GB)       & 32 & 24 & 24 & 24 \\
Board Memory Type      & DDR4 & GDDR6X & GDDR6X  &GDDR6 \\
Host CPU               & i9-10900 & Ryzen 7950X &  i9-13900K & Ryzen 7950X \\
Host \# of cores       & 10       & 16 & 24  & 16 \\
Host Memory (GB)       & 64       & 64 & 128 & 64 \\
Host Memory Type       & DDR4     & DDR5  & DDR5  & DDR5 \\
Performance of LU (Gflops)   & 7.4  & 11.8 & 12.1  & 13.4 \\  
Power Consumption (watts)   & 147  & 273     & 210   & 176  \\ 
Power Efficiency (Gflops/watts) & 0.050 & 0.043 & 0.058  & 0.076 \\
\bottomrule
\end{tabular}
\end{center}
\end{table*}

\section{Discussion} \label{sec_dis}
\subsection{Comparing FPGAs and GPUs as an accelerator for \POSITA}
Agilex FPGA and all evaluated GPUs are effective accelerators for \POSITA{}.
For example, the onboard memory size is comparable, with RTX4090, V100, and Agilex having 24, 32, and 32 GB of memory, respectively.
However, the data transfer rate to the host CPU of the GPUs is greater than that of the FPGA, so the effective performance of the FPGA is not much better than that of the GPUs, even though the FPGA is theoretically faster than the GPUs.
A critical difference between the FPGA and the GPUs is the difference in the generation of the PCI-Express interface.
Another weakness of the current design of GEMM on FPGAs is that the performance of GEMM on non-square matrices is lower than that on square matrices, as shown in Figure \ref{gemm_trailing}.

Table~\ref{taball} also compares the elapsed time for solving the two matrix decompositions between GPUs with the corresponding lowest $P_{\rm limit}$ and Agile FPGA.
For the LU decomposition, Agilex is faster than RTX3090 if we limit the power consumption of the GPU.
Note that the performance of RTX4090 and RX7900 is hardly affected by $P_{\rm limit}$. 
During the execution of the LU decomposition, the power consumption of the RX7900 board reported by the vendor API is $\sim 70$ watts.
The impact of the power limit is more severe for the LU decomposition than for the Cholesky decomposition because \Rgetrf{} calls \Rgemm{} to update the trailing matrix, which is larger than that for \Rpotrf{}.


\subsection{Comparing \POSIT{} and binary32 arithmetic on FPGAs}
As noted in section \ref{sec_acc}, Table~\ref{Agilex} shows the systolic array design using the hardware binary32 FP hardware unit of Agilex FPGA.
The design using \POSIT{}$_{\rm TC}$ requires 42\% more logic cells than the one using binary32$_{\rm Soft}$.
The number of DSP for both units is the same for both designs.
The size of the fractional part of the internal FP format of \POSIT{}$_{\rm TC}$ is 29 bits while that of the binary32 addition is 27 bits, so the size of the fractional part is almost comparable.
Because of the pre- and the post-processing, \POSIT{}$_{\rm TC}$ requires more logic cells. 

It is possible to create a much larger systolic array using the hardware FP unit.
The largest systolic array we have evaluated using the hardware FP unit is a design with $n_{\rm PE} = 96 \times 16 = 1536$, while the largest systolic array using \POSIT{}$_{\rm TC}$ is the design with 256 PEs.
This largest systolic design with the hardware FP unit consumes 34\% of all DSP blocks, so there is plenty of room for an even larger systolic array.
The measured performance of this design is $\sim 900$ Gflops with the square matrix of $10000 \times 10000$.

\subsection{Comparing \POSITA{} and binary32 arithmetic on GPUs and FPGAs}
On GPUs, the performance of GEMM in binary32 arithmetic (SGEMM) is generally close to 100\% of the theoretical peak performance.
For example, the performance of SGEMM with cuBLAS on NVIDIA A100 is close to the theoretical peak at 19.5 Gflops for square matrices with $N > 6000$ \cite{10.1177/10943420221090256}.
In their work, the authors proposed a novel method of using Tensor Cores to further accelerate SGEMM by decomposing a binary32 precision matrix into a sum of binary16/TF32 precision matrices.
Their method is more than 2 times faster than the theoretical peak of binary32 arithmetic on NVIDIA A100 and other GPUs.

The performance difference between \POSITA{} and binary32 arithmetic on GPUs and FPGAs is large because our work {\it emulates} \POSITA{} on both GPUs and FPGAs.
As shown in the previous subsection, the performance gap of GEMM in binary32 arithmetic between Agilex FPGA and the latest GPUs, such as A100 and H100, is more than a factor of 20.
Thus, the evaluation of dedicated hardware designs of POSIT arithmetic with an advanced process node size will be necessary to directly compare the performance of POSIT and binary32 arithmetic.

\section{Conclusion} \label{sec_con}
In this paper, we evaluated linear algebra accelerators in 32-bit POSIT arithmetic.
We implemented a matrix multiplication operation, a fundamental operation in linear algebra applications, on FPGAs and GPUs to accelerate two matrix decomposition algorithms.

For FPGA implementations, we have designed an accelerator of the GEMM as a systolic array architecture.
We compared two internal formats for \POSITA{} and found that the two's-complement format requires fewer hardware resources.
The performance of GEMM on our systolic-array based accelerator does not depend on the magnitude of the input matrices because, unlike the GPU implementations, the combinational logic is used to implement the pre- and post-processing of \POSITA{}.
The performance of GEMM of the Agilex FPGA is $\sim$ 200 Gflops, which is faster than that of GPUs for square matrices.
Comparing the systolic array with \POSITA{} and the standard binary32 formats on the FPGA, the POSIT-based design requires 42 \% more logic cells than the systolic array with the binary32 format. 

For GPU implementations, we first analyzed the execution of basic operations in \POSITA{} on GPUs by profiling ported GPU kernels.
We found that the performance of \POSITA{} on GPUs depends on the magnitude of variables.
When variables are close to 1.0, in a range of values called the golden zone for \POSITA{}, the performance is the fastest with the fewest number of instructions executed on a GPU.
Next, we evaluated the GEMM operations on datacenter and consumer GPUs. 
We found that the performance of GEMM also is highly dependent on the magnitude of the input matrices.
The power limit configuration is another factor that significantly degrades the performance of GEMM on RTX3090, while other GPUs are not severely affected by the power limit. 
Comparing five GPUs, RTX4090 is the fastest for GEMM with square matrices.

We analyzed the numerical error for the two decomposition algorithms.
We found that \POSITA{} is more effective when the norm of the elements of the input matrix is in the golden zone.
For the Cholesky and LU decompositions, the relative backward errors of \POSITA{} are 0.5 and 0.8 digits more accurate than those of binary32 arithmetic, respectively. 

Finally, FPGA and GPU systems effectively accelerate the Cholesky and LU decomposition algorithms.
However, the data transfer rate of FPGAs to/from the host CPU is slower than that of the GPUs, so the consumer GPUs are faster regarding the matrix decomposition algorithms. 
By limiting the power consumption of the consumer GPU to 100-150 watts, we found that RTX3900 is slower than Agilex in GEMM and matrix decompositions.
At the same time, the performance of RTX4090 and RX7900 is barely affected by the power limit.
By measuring the AC power of FPGA and GPU systems, we evaluated the power efficiency \POSITA{} for solving the LU decomposition.
The power efficiency of the FPGA as an accelerator is 0.05 Gflops/watts, while that of RX7900 GPU is 0.076 Gflops/watts. 
Note that the measurement of the power consumption depends on many conditions such as the size of the input matrix, a host CPU, and a power supply unit of a system.

As reported in \cite{2023arXiv230611975O}, the effective use of the integer operations on recent GPUs is a hot topic in HPC and ML research. 
Our result also shows that recent GPUs are complementary platforms to FPGAs for evaluating \POSITA{} by using the emulation with integer operations.
An extension of our work to other arithmetic, such as IEEE binary128 and shorter and longer data length arithmetic formats, will be our future work.




\begin{acks}
This work is partly supported by MEXT as ``Feasibility studies for the next-generation computing infrastructure'' and KAKENHI Grant Number JP23K11133.
In a part of this work, we used computational resources of Cygnus provided by Multidisciplinary Cooperative Research Program in Center for Computational Sciences, University of Tsukuba.
\end{acks}

\bibliographystyle{ACM-Reference-Format}
\bibliography{main}

\end{document}